\documentclass[aps,prc,showpacs,floatfix,footinbib]{revtex4}

\usepackage{subfigure}
\usepackage{amsmath}    
\usepackage{graphicx}   
\usepackage{verbatim}  
\usepackage{color}      
\usepackage{hyperref}

\begin{document}

\title{Thermodynamics and fluctuations of conserved charges in Hadron Resonance Gas model in finite volume}
\author{Abhijit Bhattacharyya}
\email{abphy@caluniv.ac.in}
\affiliation{Department of Physics, University of Calcutta,
92, A. P. C. Road, Kolkata - 700009, INDIA}
\author{Rajarshi Ray}
\email{rajarshi@jcbose.ac.in}
\author{Subhasis Samanta}
\email{samanta@jcbose.ac.in}
\affiliation{Center for Astroparticle Physics \&
Space Science, Bose Institute, Block-EN, Sector-V, Salt Lake, Kolkata-700091, INDIA 
 \\ \& \\ 
Department of Physics, Bose Institute, \\
93/1, A. P. C Road, Kolkata - 700009, INDIA}
\author{Subrata Sur}
\email{ssur.phys@gmail.com}
\affiliation{Department of Physics, Panihati Mahavidyalaya,
Barasat Road, Sodepur, Kolkata - 700110, INDIA}

\begin{abstract}
The thermodynamics of hot and dense matter created in heavy-ion
collision experiments are usually studied as a system of infinite
volume. Here we report on possible effects for considering a finite
system size for such matter in the framework of the Hadron Resonance Gas
model. The bulk thermodynamic variables as well as the fluctuations of
conserved charges are considered. We find that the finite size effects
are insignificant once the observables are scaled with the respective
volumes. The only substantial effect is found in the fluctuations of
electric charge which may therefore be used to extract information
about the volume of fireball created in heavy-ion collision experiments.
\end{abstract}

\pacs{12.38.Mh, 21.60.-n,21.65.Mn,25.75.-q}

\maketitle

Our present day universe contains a significant fraction of matter in
hadronic form. In the very early universe $-$ a few microseconds after
the Big Bang~\cite{kolb_turner} when the temperature was extremely high,
the strongly interacting matter is expected to have existed in the
partonic form. Similar exotic state of matter may exist inside compact
stars due to extremely high matter density attained by gravitational
compression~\cite{wilczek}. For the last few decades various
experimental efforts are being made to recreate such exotic matter
through the collisions of heavy-ions at ultra-relativistic energies.
Experimental facilities at CERN (France/Switzerland), BNL (USA) and the
upcoming facility at GSI (Germany) are at the forefront of efforts
taken to create these exotic states of matter. One of the major goals in
the experiments is to study thermodynamic properties of strongly
interacting matter at high temperatures and densities. Present
experimental data as well as lattice QCD simulations seem to indicate a
smooth cross over from hadronic to quark gluon matter at low density and
high temperature
\cite{nature05120_Aoki, PRL65_2491_Brown}. At high density and low
temperature a first order transition is expected
\cite{PRD78_074507_Ejiri, PRD29_Pisarski, NPA504_Asakawa,
PRD58_096007_Halasz, PRD67_014028_Hatta, PRC79_Bowman}.

Usually any thermodynamic study assumes the system volume to be
infinite. However the fireball created in the relativistic heavy ion
collision experiments has a finite spatial volume. The size of such
spatial volume critically depends on three parameters : the size of the
colliding nuclei, the center of mass energy ($\sqrt{s}$) and the
centrality of collisions. Analysis of experimental data could reveal the
freeze out volume of the system. One way to carry out such an analysis
is the study of HBT radii which has been done in Ref. \cite{Adamova}.
The major finding of this study indicates that the freeze out volume
increases as the $\sqrt{s}$ increases and the estimated freeze out
volume varies from 2000 $fm^3$ to 3000 $fm^3$. Another way of estimating
the system size is through the comparison of simulation results with the
experimental data as done in Ref. \cite{Graef} where the UrQMD
model \cite{Bass} is used for this purpose. A study of the $Pb-Pb$
collisions at different energies and centralities resulted in a
freeze-out volume in the range 50 $fm^3$ to 250 $fm^3$. We note that
these volumes as quoted above, are the freeze out volumes and as one
looks back to the early times in the evolution of the system, a
smaller volume is expected. So both in the quark phase and in
the hadron phase (after the putative phase transition) finite volume
could play an important role in the quantitative measurement of the
observables. Thus a thorough study of finite volume effects on the
thermodynamic observables in strongly interacting matter at high
temperature and density is required. Some aspects of such effects
have been discussed in the literature in the context of QCD inspired
models as well as pure gluon theory on the lattice \cite{fss,fss1,fss2,
fss3,elze,spieles,fslat1,fslat2,fischer1,fischer2,lusher1,gasser1,
kiriyama,shao,braun1,braun2,khanna,yasui,abreu1,abreu2,ebert,ab,sur}.
These studies indicate a significant effect of finite volume both in
the vacuum and also in the medium. Particularly, in the medium the
studies in finite volumes have a strong bearing on the location of the
critical temperature, critical end point, and related thermodynamic
variables as discussed by some of us in \cite{ab,sur}. Even the
existence of critical end point, which in turn depends on the order of
the phase transition, depends on the volume \cite{ab}. In \cite{sur} we
found that for a two flavour system the scaling of ratios of fluctuations
with volume is violated around the phase boundary.

Most of the studies mentioned above are related to the partonic degrees
of freedom. However the matter detected in the experiments are in the
hadronic form and looking at the freeze out volumes one can easily
conclude that the finite volume effects would be relevant here. So, a
finite volume study of the hadronic matter at finite temperature and
density would be useful. To this end we take recourse to Hadron
Resonance Gas (HRG) model \cite{PLB97_Hagedorn}. HRG model is based on
the Dashen, Ma and Bernstein theorem~\cite{dashen} which shows that a
dilute system of strongly interacting matter can be described by a gas
of free resonances. The attractive part of the hadron interactions is
supposed to be taken care of by these resonances. 

In this work we plan to study hadronic matter at high temperature and
density confined to a finite volume. We shall study the thermodynamic
properties of hadronic matter and also the fluctuations of various
conserved charges. We have organised the paper as follows. First the
Hadron Resonance Gas model and its finite volume extension is
introduced. We then discuss the method for calculating fluctuation of
different conserved charges. Thereafter we present our results and
conclude.

In the existing literature one may find a variety of HRG models.
Detailed discussions of the different versions of HRG model and
some of the recent works using these models may be found in 
Refs.~\cite{PLB97_Hagedorn, HRG_Braun-Munzinger, PLB344_Braun-Munzinge,
arXiv:nucl-th/9603004_Cleymans, PLB465_Braun-Munzinge,
PRC60_054908_Cleymans, PRC73_Becattini, PLB518_Braun-Munzinger,
NPA772_Andronic, PLB673_Andronic, ZPC51_Rischke, PS48_277_Cleymans,
 Singh, PRC56_Yen, PLB718_Andronic,mohanty,samanta}.
In this work we plan to discuss the simplest version namely the
non-interacting HRG. The basic physics should be independent of
which version of the model we choose. First we briefly describe the HRG
model. 

The grand canonical partition function of a hadron resonance 
gas~\cite{HRG_Braun-Munzinger, PLB718_Andronic} can be written as,
\begin {equation}
 \ln Z^{id}=\sum_i \ln Z_i^{id},
\end{equation}
where sum is over all the hadrons, $id$ refers to ideal {\it i.e.},
non-interacting HRG. For the $i$'th species,
\begin{equation}
 \ln Z_i^{id}=\pm \frac{Vg_i}{2\pi^2}\int_0^\infty p^2\,dp
 \ln[1\pm\exp(-(E_i-\mu_i)/T)],
\end{equation}
where $E_i=\sqrt{{p}^2+m^2_i}$ is the single particle energy, $m_i$ is
the mass, $V$ is the volume of the system, $g_i$ is the degeneracy
factor and $T$ is the temperature. In the above expression
$\mu_i=B_i\mu_B+S_i\mu_S+Q_i\mu_Q$ is the chemical potential and
$B_i,S_i,Q_i$ are respectively the baryon number, strangeness and charge
of the particle, $\mu^,s$ being corresponding chemical potentials. The
$(+)$ and $(-)$ sign corresponds to fermions and bosons respectively.

Our purpose is to consider HRG in a system of finite volume. We
incorporate this by considering a lower momentum cut-off
$p_{min}=\pi/R=\lambda$(say) where $R$ is the size of a cubic volume.
With this cut-off the partition function for particle $i$ becomes
\begin{equation}\label{eq:zi}
 \ln Z_i^{id}=\pm \frac{Vg_i}{2\pi^2}\int_\lambda^\infty p^2\,dp
 \ln[1\pm\exp(-(E_i-\mu_i)/T)],
\end{equation}
From partition function we can calculate various thermodynamic
quantities of interest. The partial pressure $P_i$,
the energy density $\varepsilon_i$ and the entropy density $s_i$
can be calculated using the standard definitions,

\begin{align}\label{eq:p}
 \begin{split}
  P_i^{id}=T\frac{\partial \ln Z_i^{id}}{\partial V},
 \end{split}
\end{align}

\begin{align}
\begin{split}
\varepsilon_i^{id}&=\frac{E_i^{id}}{V}=
 -\frac{1}{V} \left(\frac{\partial \ln Z_i^{id}}
                         {\partial\frac{1}{T}}\right)_{\frac{\mu}{T}}
=\frac{g_i}{2\pi^2}\int_\lambda^\infty
 \frac{p^2\,dp}{\exp[(E_i-\mu_i)/T]\pm1}E_i,
\end{split}
\end{align}

\begin{align}\label{eq:s}
 \begin{split}
  s_i^{id}&=\frac{S_i^{id}}{V}=
\frac{1}{V}\left(\frac{\partial\left({T \ln Z_i^{id}}\right)}
                                     {\partial T}\right)_{V,\mu}
 = \pm\frac{g_i}{2\pi^2}\int_\lambda^\infty p^2\,dp
        \left[ \ln\left(1\pm\exp(-\frac{(E_i-\mu_i)}{T})\right)\right.
 \left.\pm\frac{(E_i-\mu_i)}{T(\exp((E_i-\mu_i)/T)\pm1)}\right].
 \end{split}
\end{align}
The specific heat at constant volume $C_V$ is given by,

\begin{align}\label{eq:cv}
 \begin{split}
  C_V^{id}&=\left(\frac{\partial E_i^{id}}{\partial T}\right)_V
  =\frac{g_iV}{2\pi^2}\int_\lambda^\infty
\frac{p^2\,dp\exp((E_i-\mu_i)/T)}
     {T^2(\exp((E_i-\mu_i)/T)\pm1)^2}(E_i-\mu_i)E_i.
 \end{split}
\end{align}

Here we consider the grand canonical partition function given in
equation (\ref{eq:zi}) to incorporate the effect of finite volume and
have incorporated all the hadrons listed in the particle data book
\cite{PDG} up to mass of $3$ GeV.
 
\begin{figure}[!hbt]
\centering
 \subfigure {\includegraphics[scale=0.4]{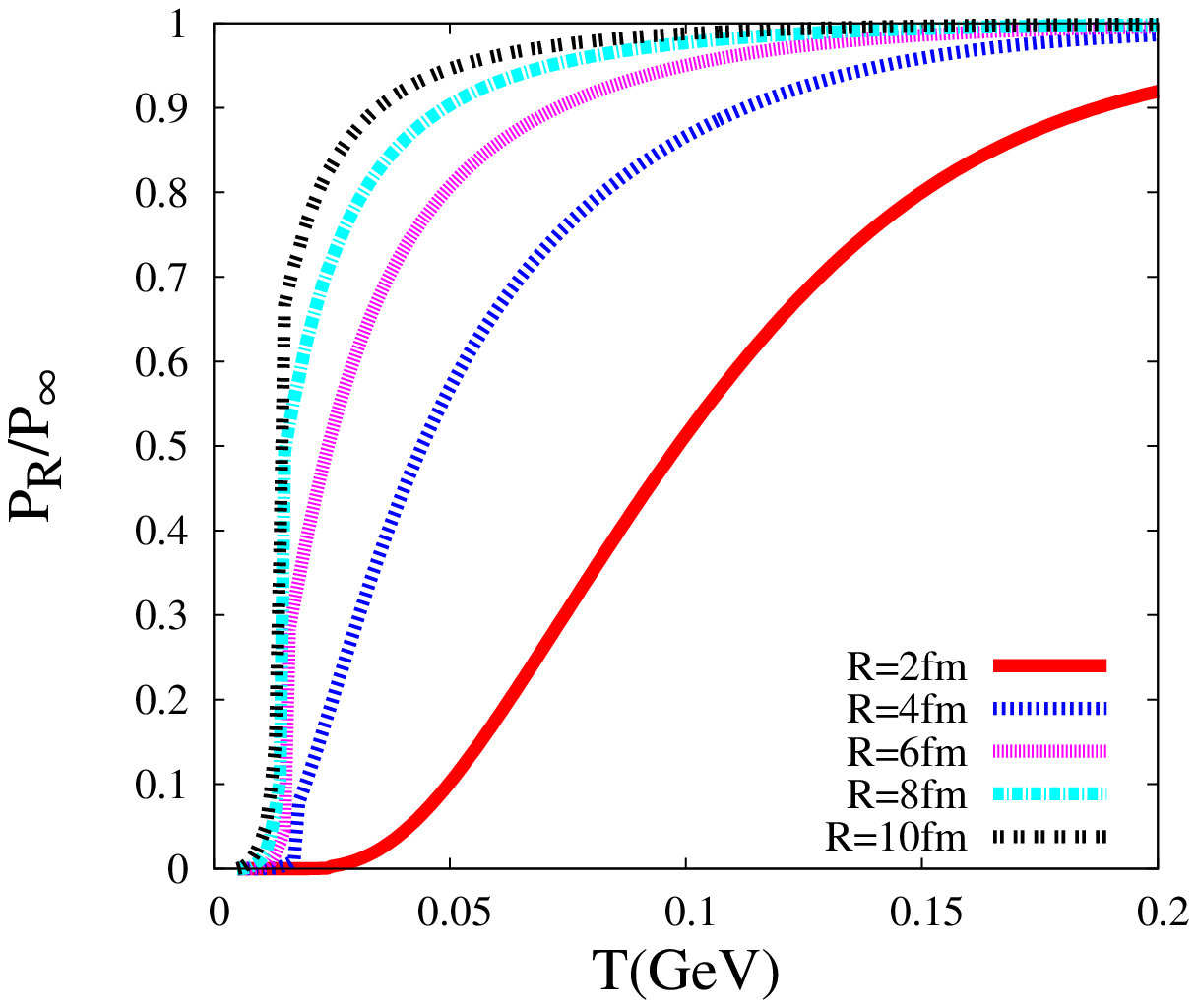}
             \label {press_temp}}
 \subfigure {\includegraphics[scale=0.4]{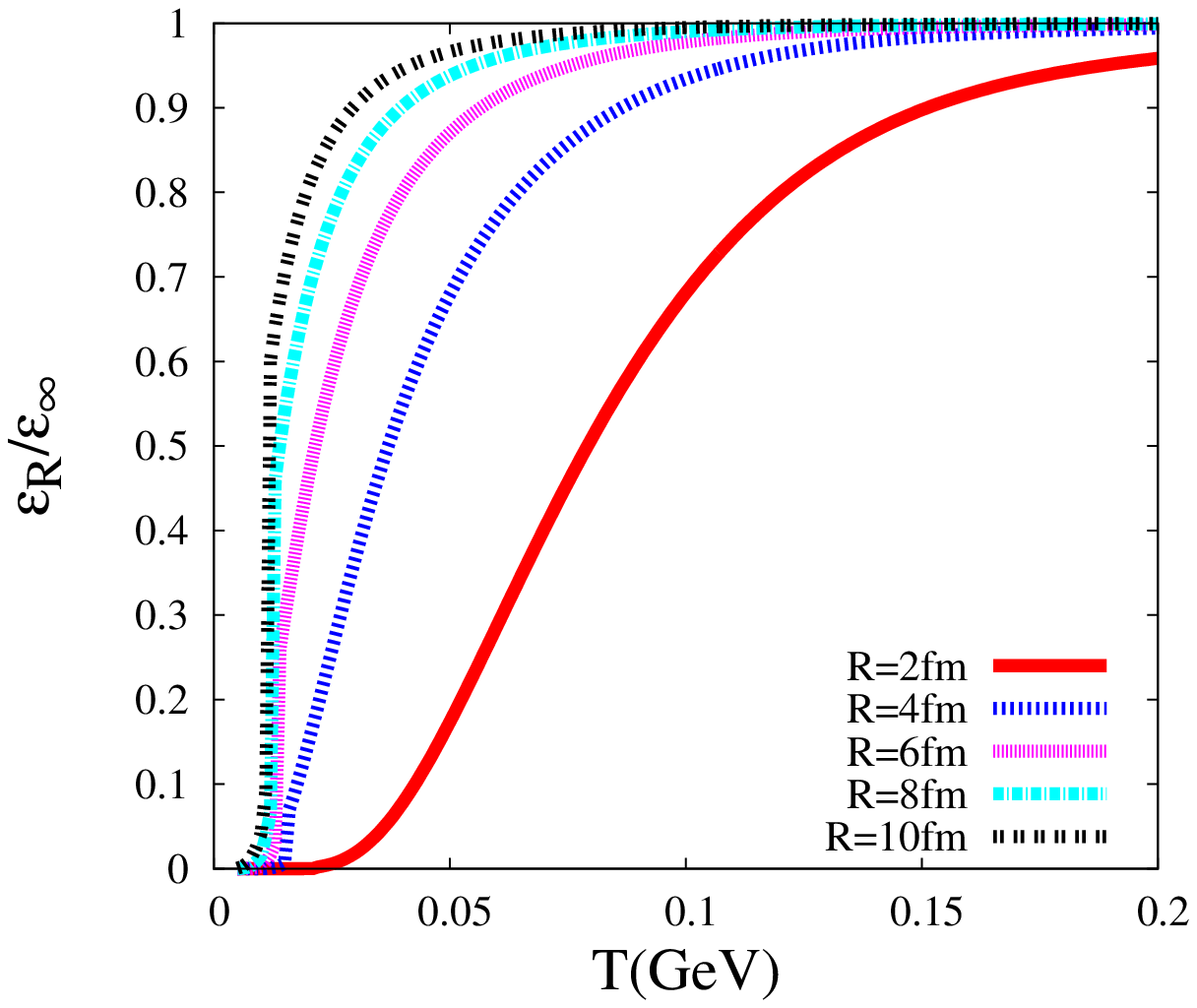}
             \label {energy_temp}}
 \subfigure {\includegraphics[scale=0.4]{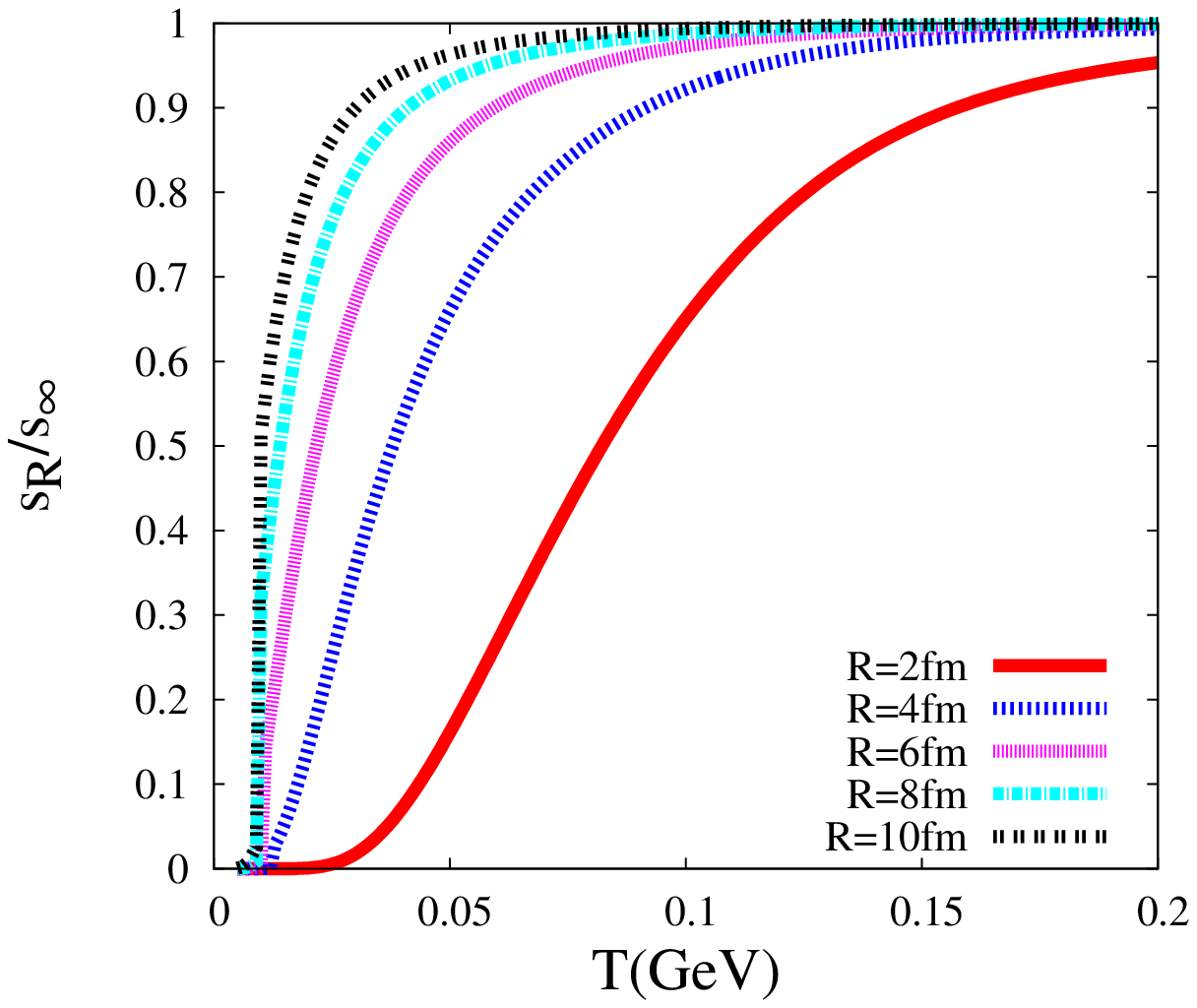}
             \label {entropy_temp}}
 \subfigure {\includegraphics[scale=0.4]{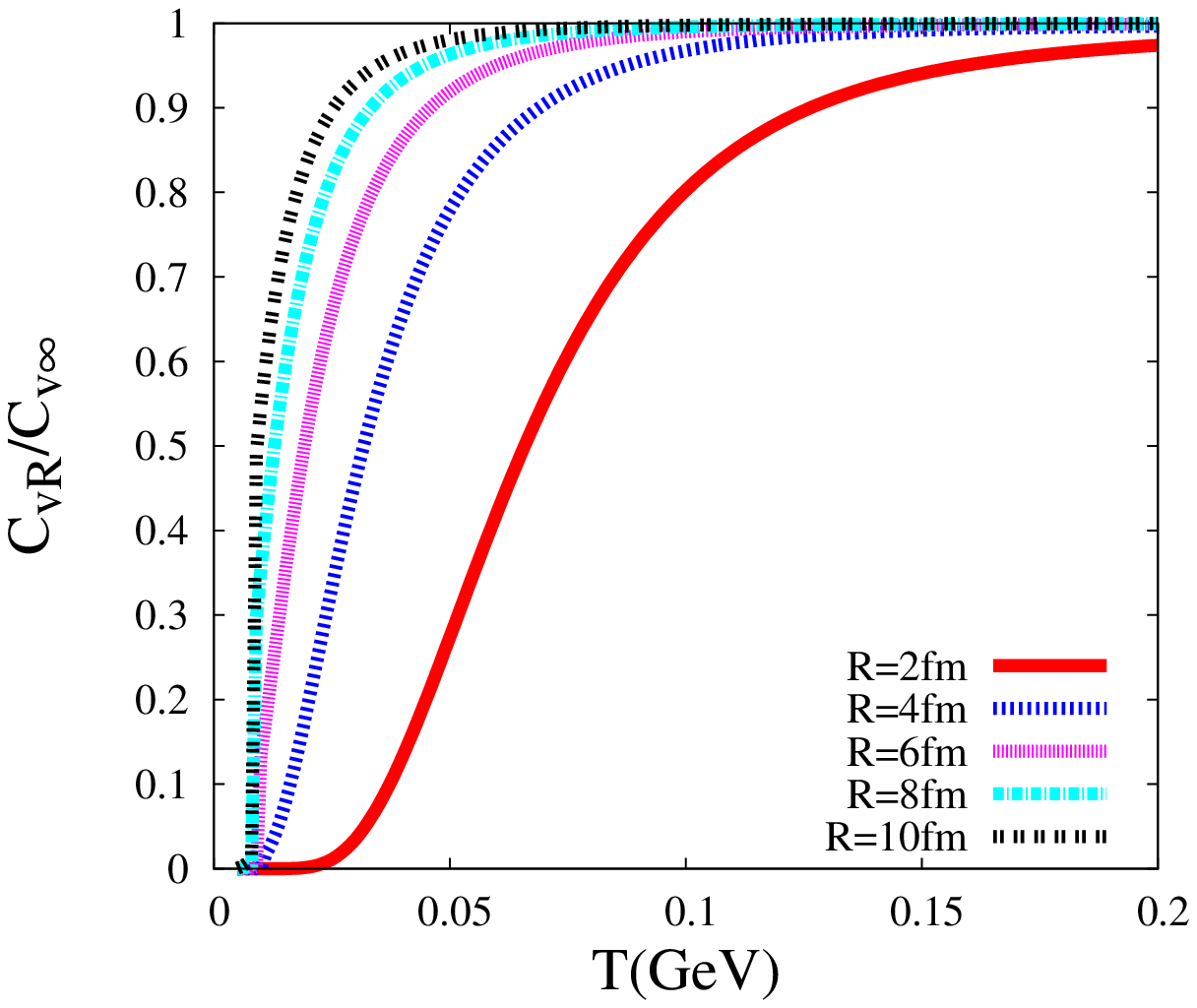}
             \label {cv_temp}}
 \caption{(Color online) Variation of scaled pressure, scaled energy,
           scaled entropy and scaled specific heat with temperature. } 
 \label{fig:eos_temp}
\end{figure}

In Fig. \ref{fig:eos_temp} we have shown the scaled pressure {\it i.e.},
the ratio of pressure calculated for a finite system size $R$ to the
pressure for infinite volume ( {\it i.e.} $R= \infty$) as a function of
temperature $T$. Also shown in the adjacent figures are scaled energy,
scaled entropy and scaled specific heat. Five different representative
system sizes are chosen: $R=2fm$, $R=4fm$, $R=6fm$, $R=8fm$ and
$R=10fm$. We find a significant volume dependence in all these
quantities. At extremely low temperature ($\sim$ 0.02 GeV) the finite
volume effect is strongest. As the temperature increases the scaled
variables approach unity. This is expected since the lowest lying
hadrons are most dominant at low temperatures and they feel the finite
volume effects the most. With increase in temperature higher mass
resonances become important and the dependence on finite system size
diminishes.

 At around a temperature of $0.1$ GeV the scaled quantities for
$R=10fm$ almost reaches one indicating that the results are almost
indistinguishable from those at infinite volume. Obviously for smaller
volumes the scaled variables will reach unity at higher temperatures. 
We find that by $0.2$ GeV of temperature the scaled quantities for
system sizes down to $R=4fm$ reach the value of one. For smaller
volumes the system will still be away from the infinite volume scenario
at $T = 0.2$ GeV by which the system is supposed to convert to a
partonic phase.

Fluctuations of conserved charges such as net electric charge, baryon
number and strangeness have been considered as probes for hadronization
and thermalization of the system created in nuclear collisions
\cite{Asakawa, Bower, Aziz}. Moreover, fluctuations are expected to
show distinctly different behaviour in a hadron resonance gas and a QGP. 
From the grand canonical partition function ($Z$) we take derivatives
with respect to chemical potential to get the various susceptibilities.
The $n^{th}$ order susceptibility is defined as,
\begin{equation}\label{eq:chi}
 \chi^n_x=\frac{1}{V T^3}
\frac{\partial^n {(ln Z)}}{\partial {(\frac{\mu_x}{T})}^n},
\end{equation}
where $\mu_x$ is the chemical potential for conserved charge $x$. We
have considered for our purpose $x=B$ (baryon), $S$ (strangeness) and
$Q$ (electric charge).

\begin{figure}[!htb]
\centering
  \subfigure {\includegraphics[scale=0.4]{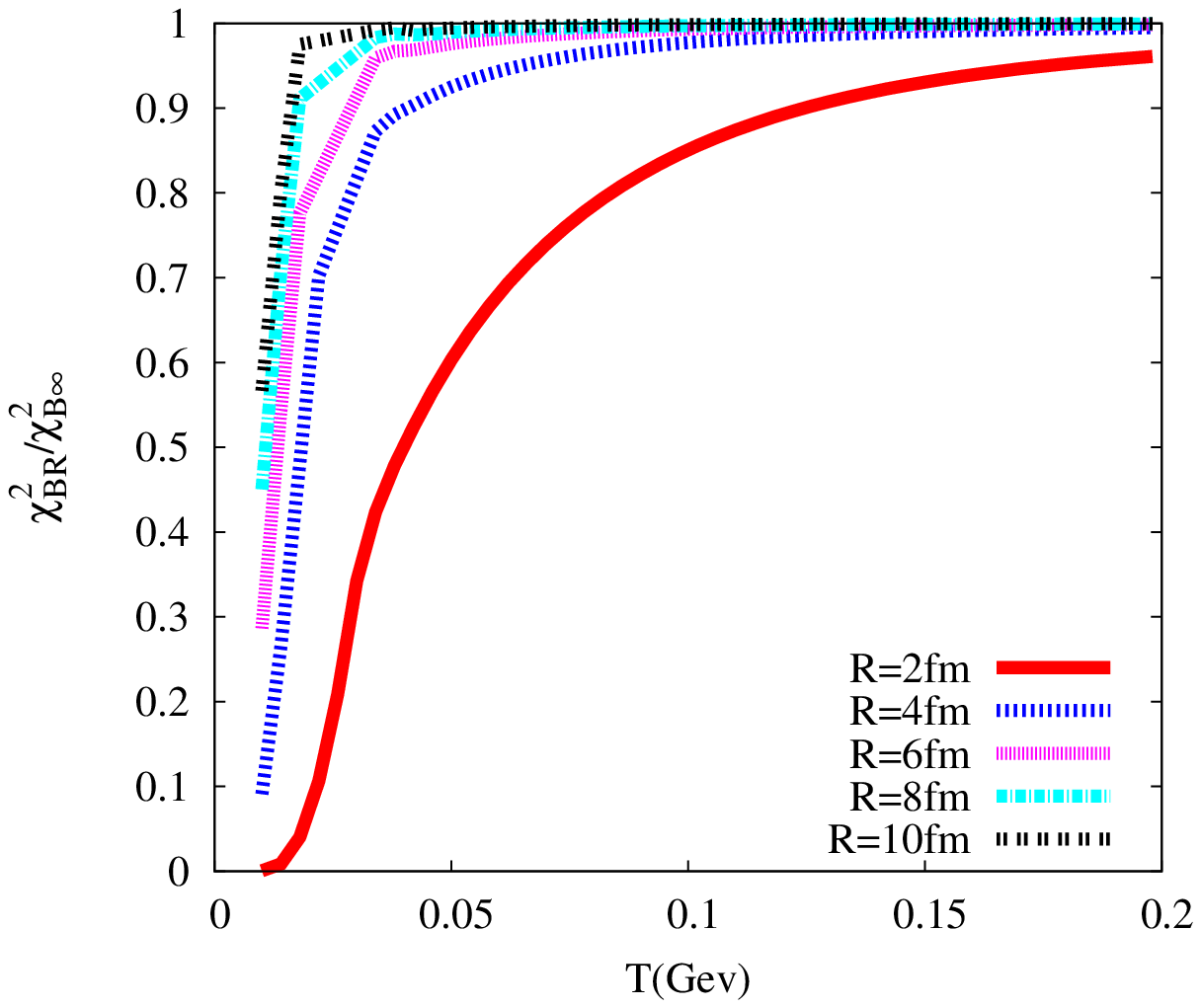}
              \label {chi_B2_temp}}
  \subfigure {\includegraphics[scale=0.4]{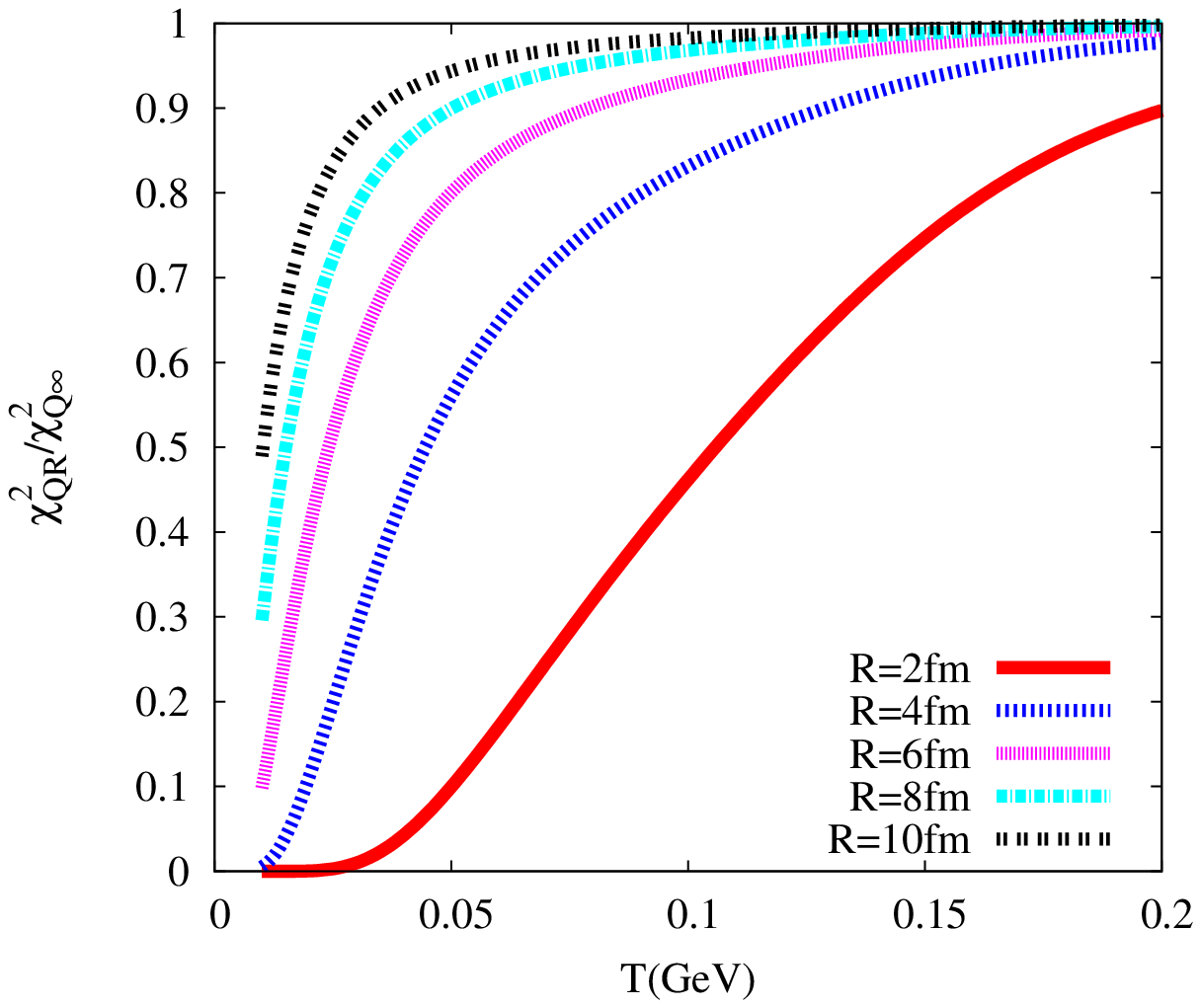}
              \label {chi_Q2_temp}}
  \subfigure {\includegraphics[scale=0.4]{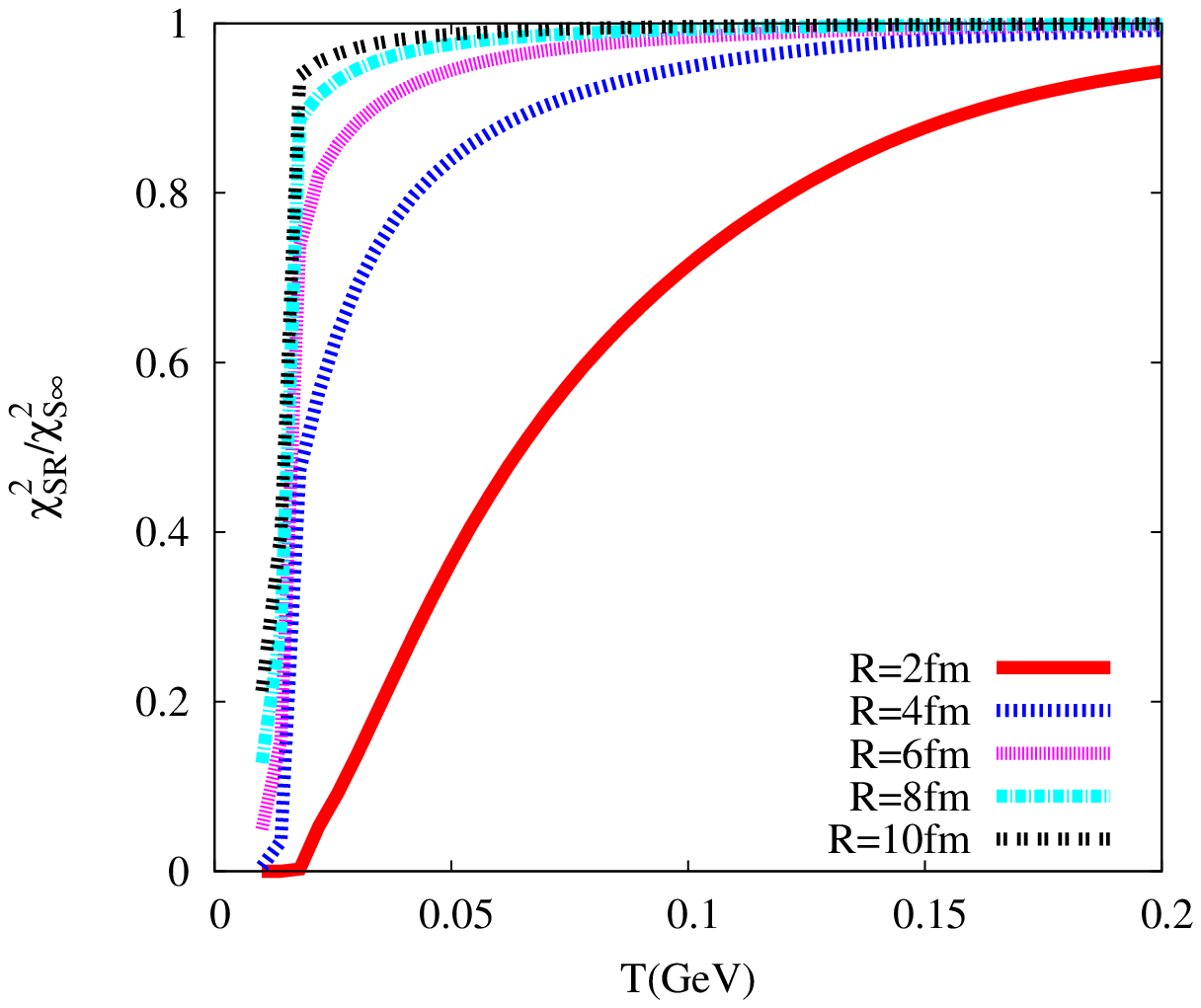}
              \label {chi_S2_temp}}
  \subfigure {\includegraphics[scale=0.4]{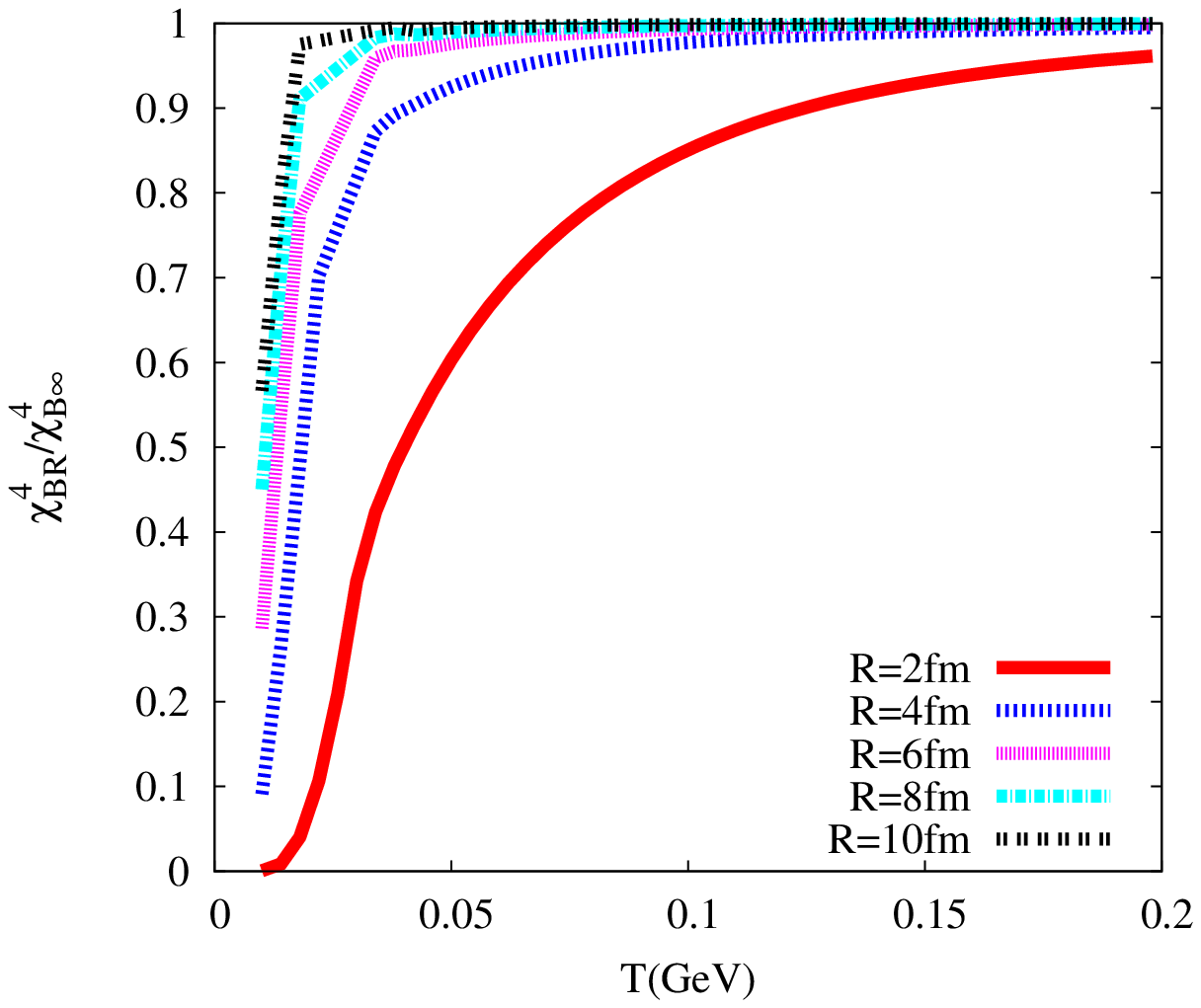}
              \label {chi_B4_temp}}
  \subfigure {\includegraphics[scale=0.4]{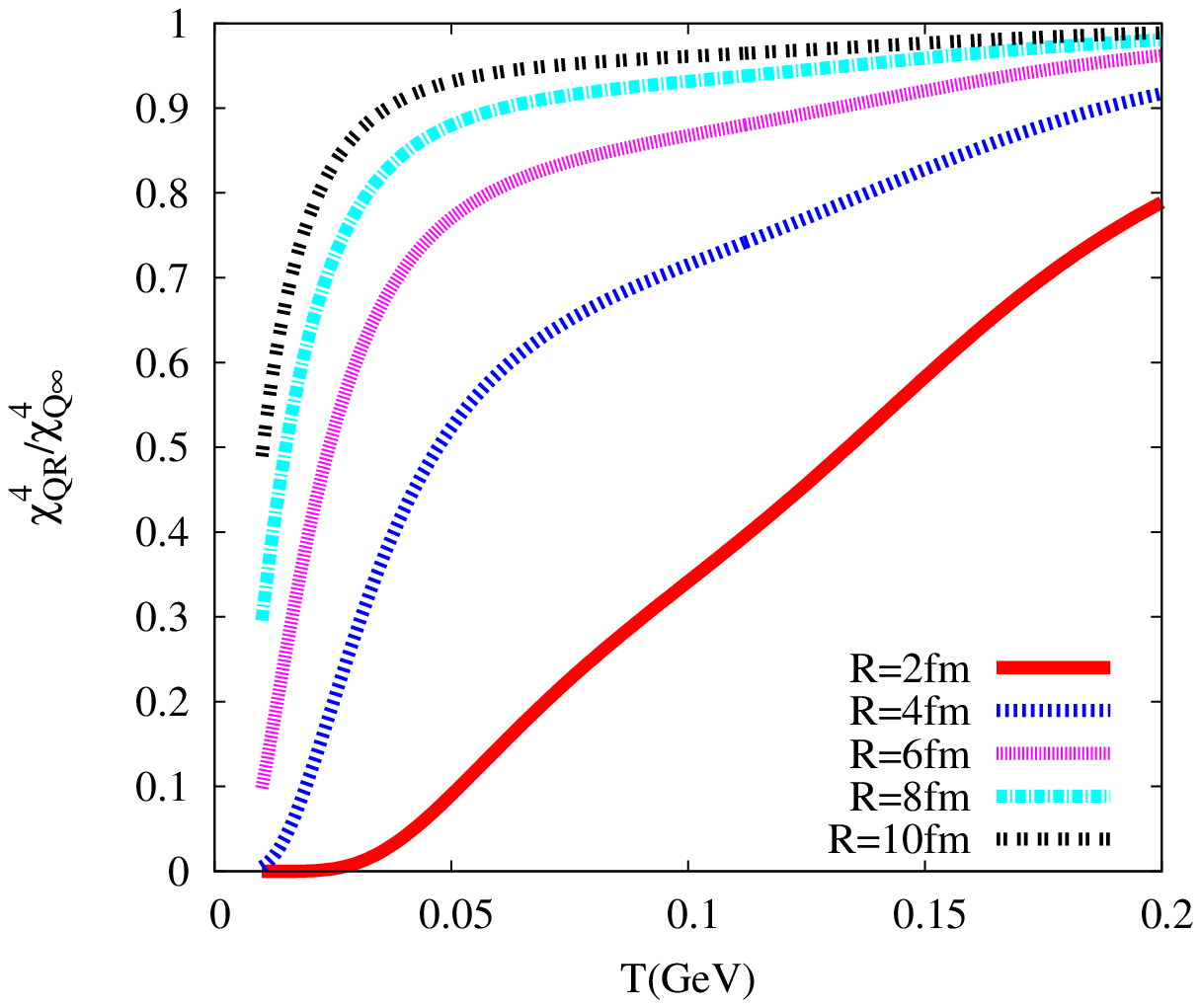}
              \label {chi_Q4_temp}}
  \subfigure {\includegraphics[scale=0.4]{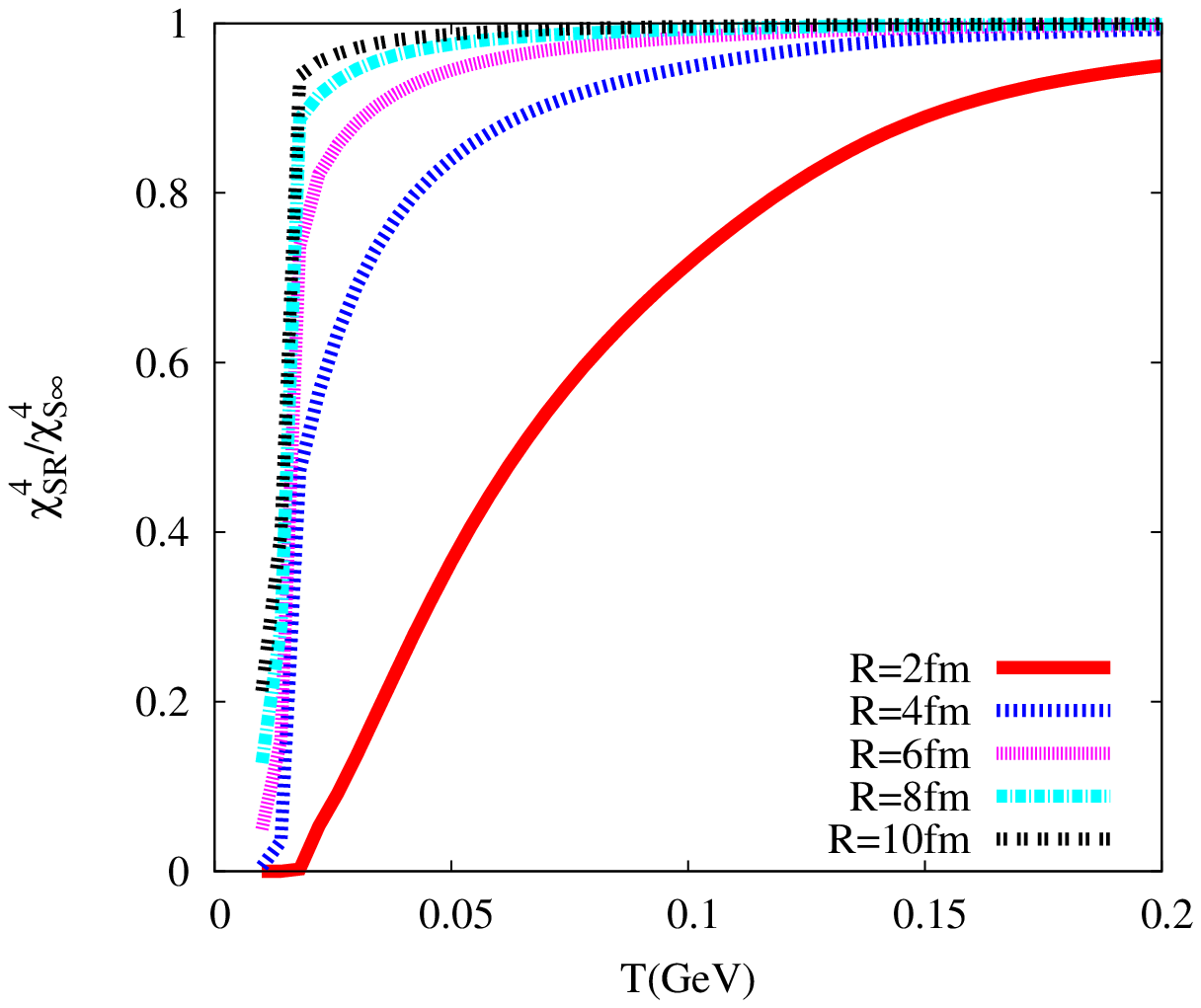}
              \label {chi_S4_temp}}   
 \caption{(Color online) Variation of scaled susceptibilities $\chi^2$,
 $\chi^4$, for baryon (left column), electric charge (middle column) and strangeness 
 (right column), with temperature ($\mu=0$).}
 \label{fig:chiB_temp}
\end{figure}

In Fig. \ref{fig:chiB_temp} we have shown scaled second order and fourth
order susceptibilities for conserved charges namely baryon number,
electric charge and strangeness as function of temperature for $\mu_B=0$
for different system sizes $R$. The general features of these scaled
fluctuations are similar to the previous thermodynamic variables. For
all the cases we observe a strong volume dependence especially for low
temperature. Though the quantitative features of baryon number and
strangeness are very close to those of the thermodynamic variables
discussed earlier, the nature of electric charge fluctuations are quite
different. We observe that in this case even the systems with volumes
larger than 4 $fm$ are significantly different from the infinite volume
systems. It may be noted that the lightest hadrons - the pions
contribute to this sector and not to the other charge fluctuations. As
expected the ensemble with the lightest particles will have the most
significant signatures of finite volumes.

\begin{figure}[!htb]
   \subfigure {\includegraphics[scale=0.4]{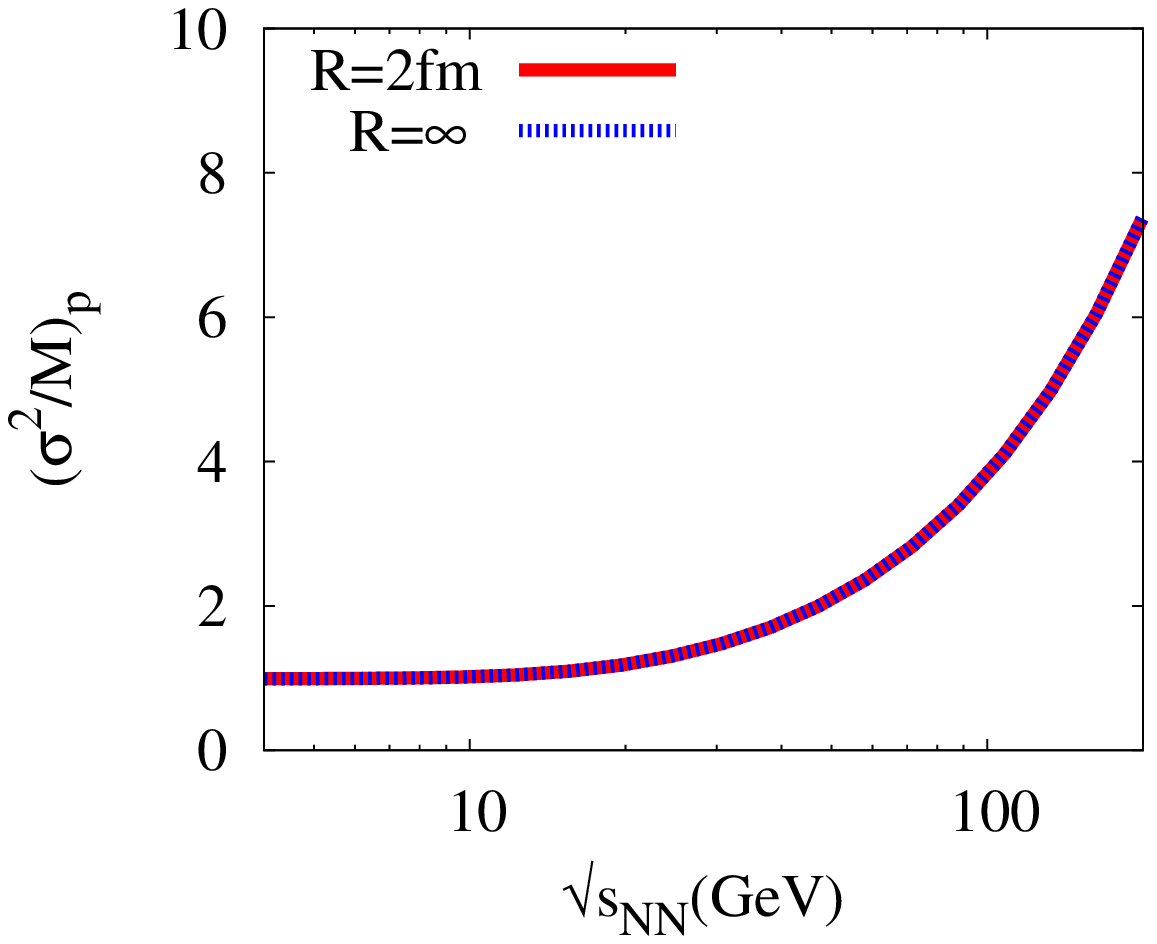}
               \label {s2_np_roots}}
   \subfigure {\includegraphics[scale=0.4]{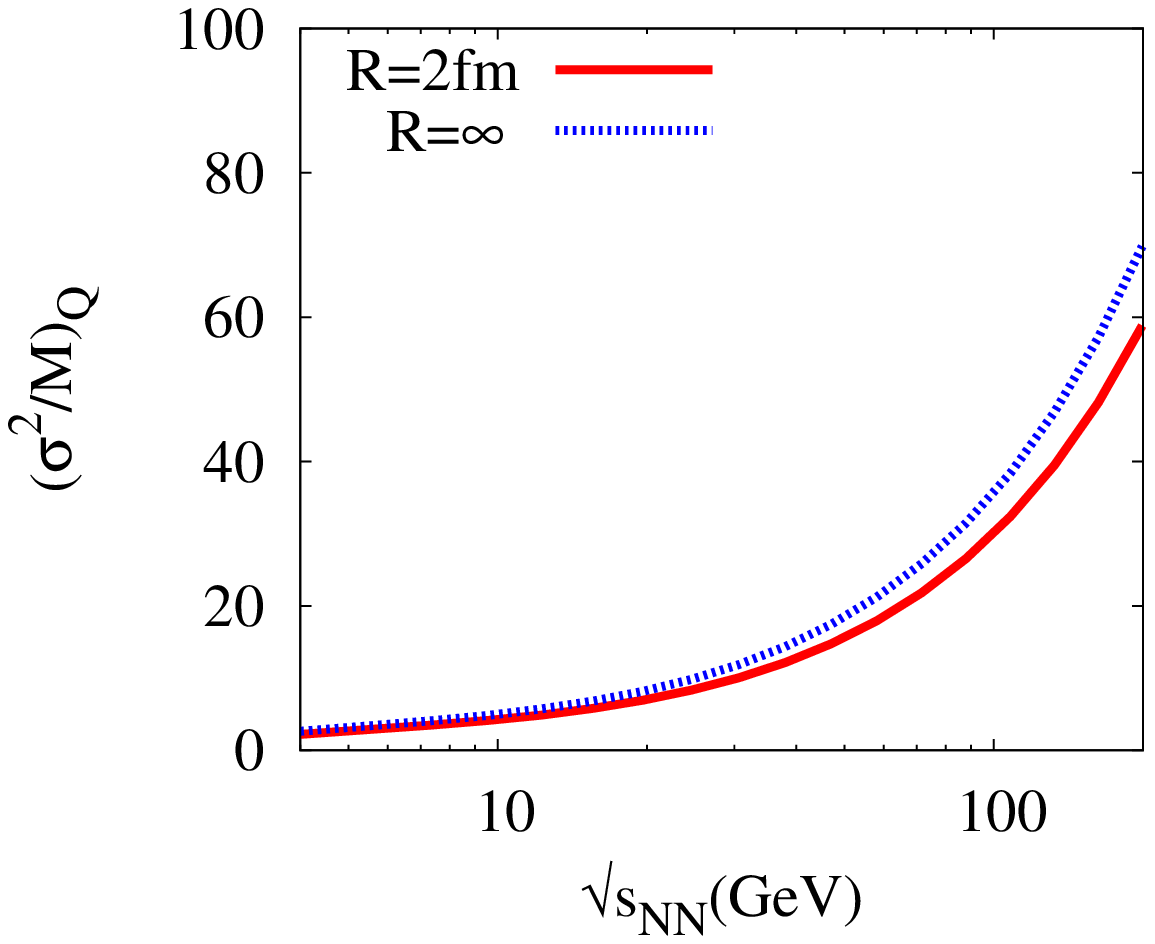}
               \label {s2_nc_roots}}
   \subfigure {\includegraphics[scale=0.4]{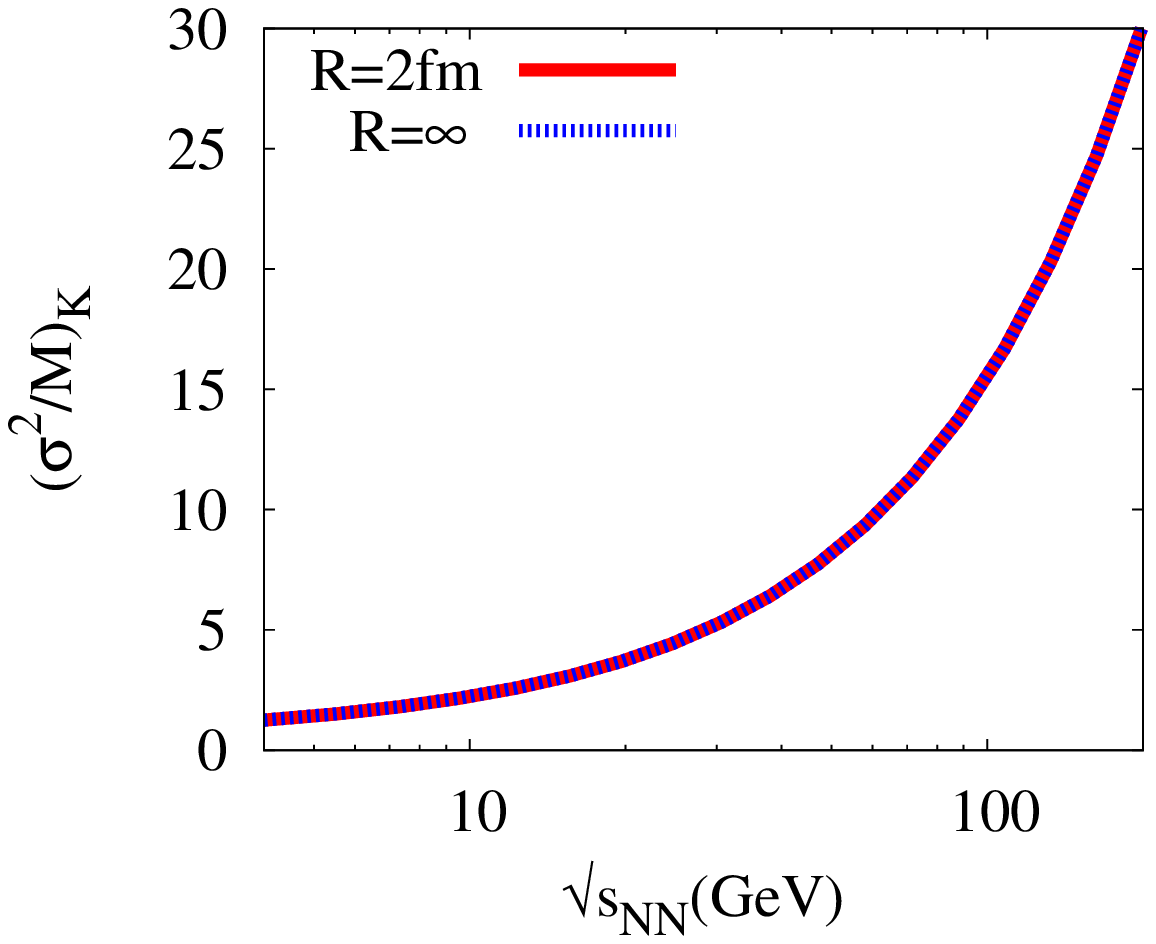}
               \label {s2_nk_roots}}
   \subfigure {\includegraphics[scale=0.4]{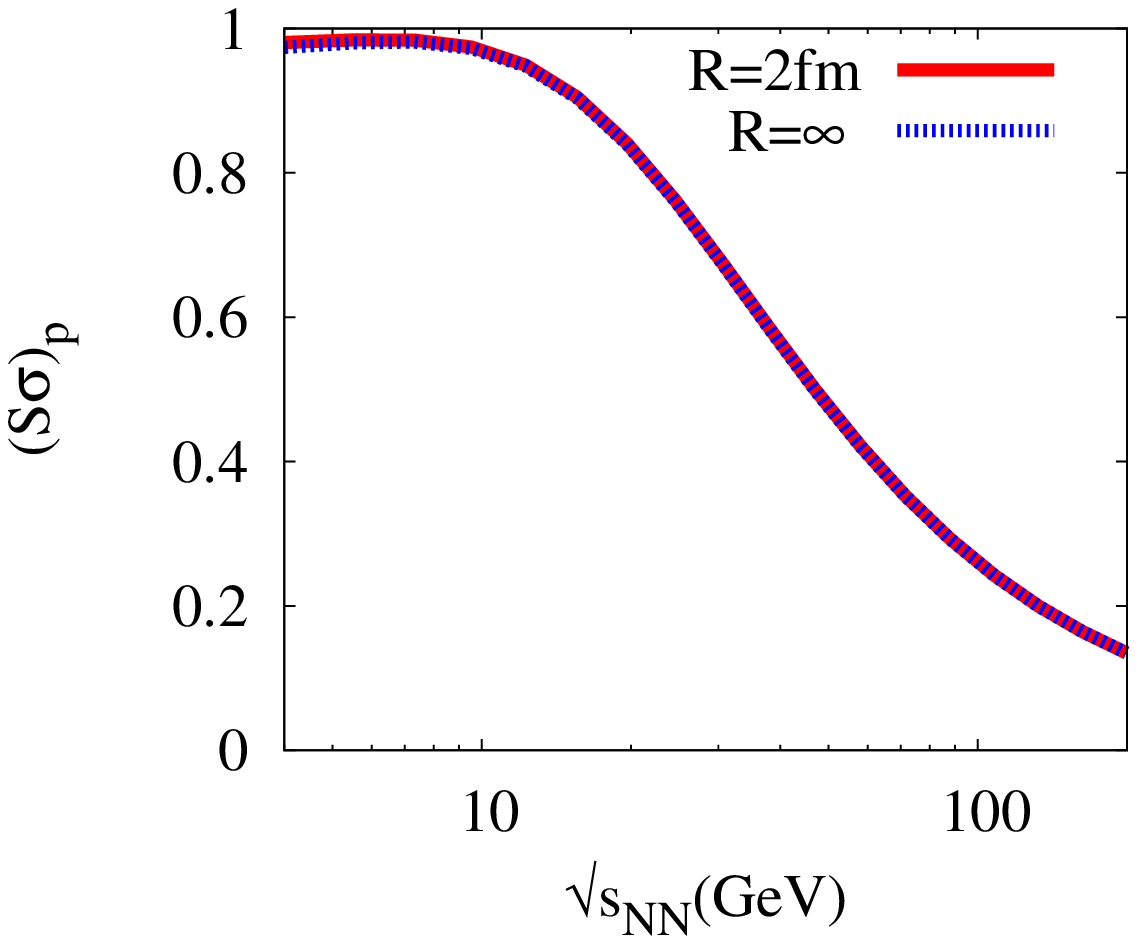}
               \label {ss_np_roots}}
   \subfigure {\includegraphics[scale=0.4]{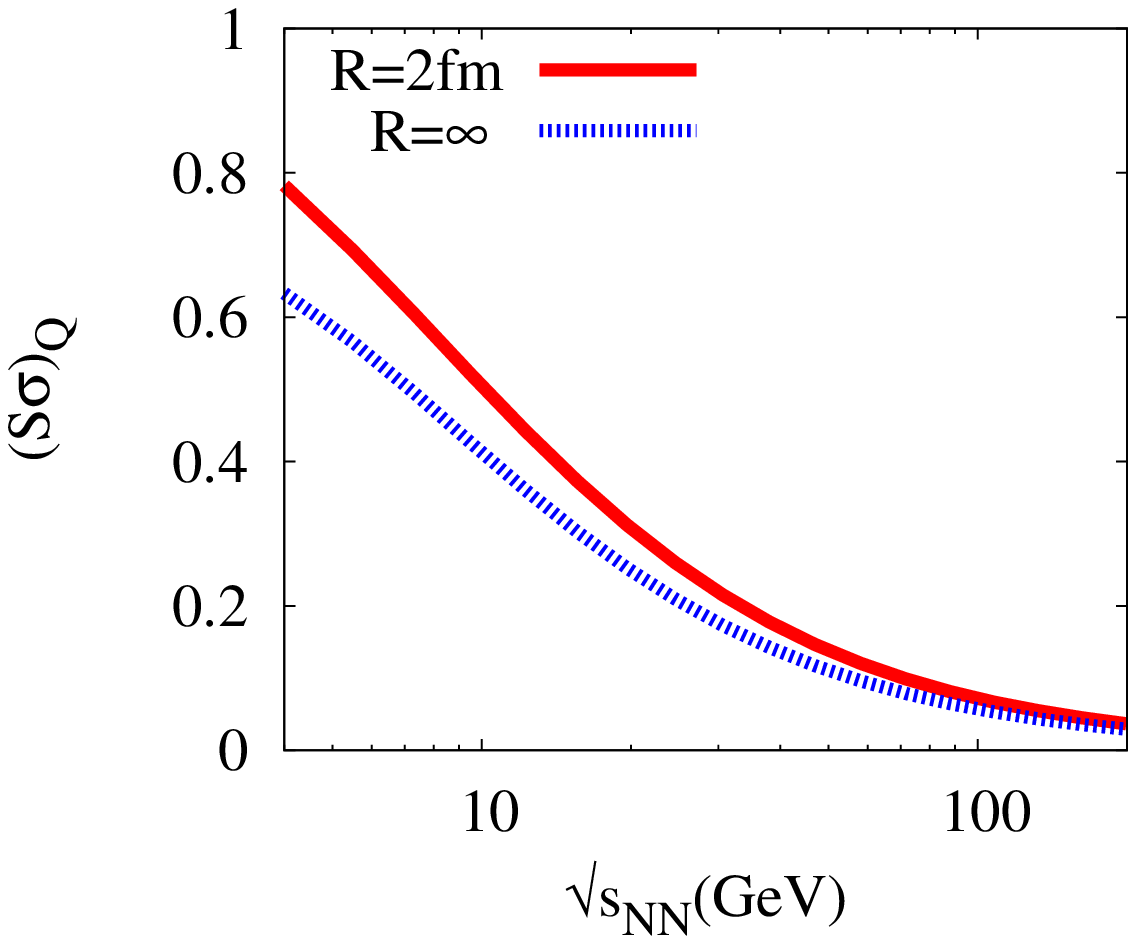}
               \label {ss_nc_roots}}
   \subfigure {\includegraphics[scale=0.4]{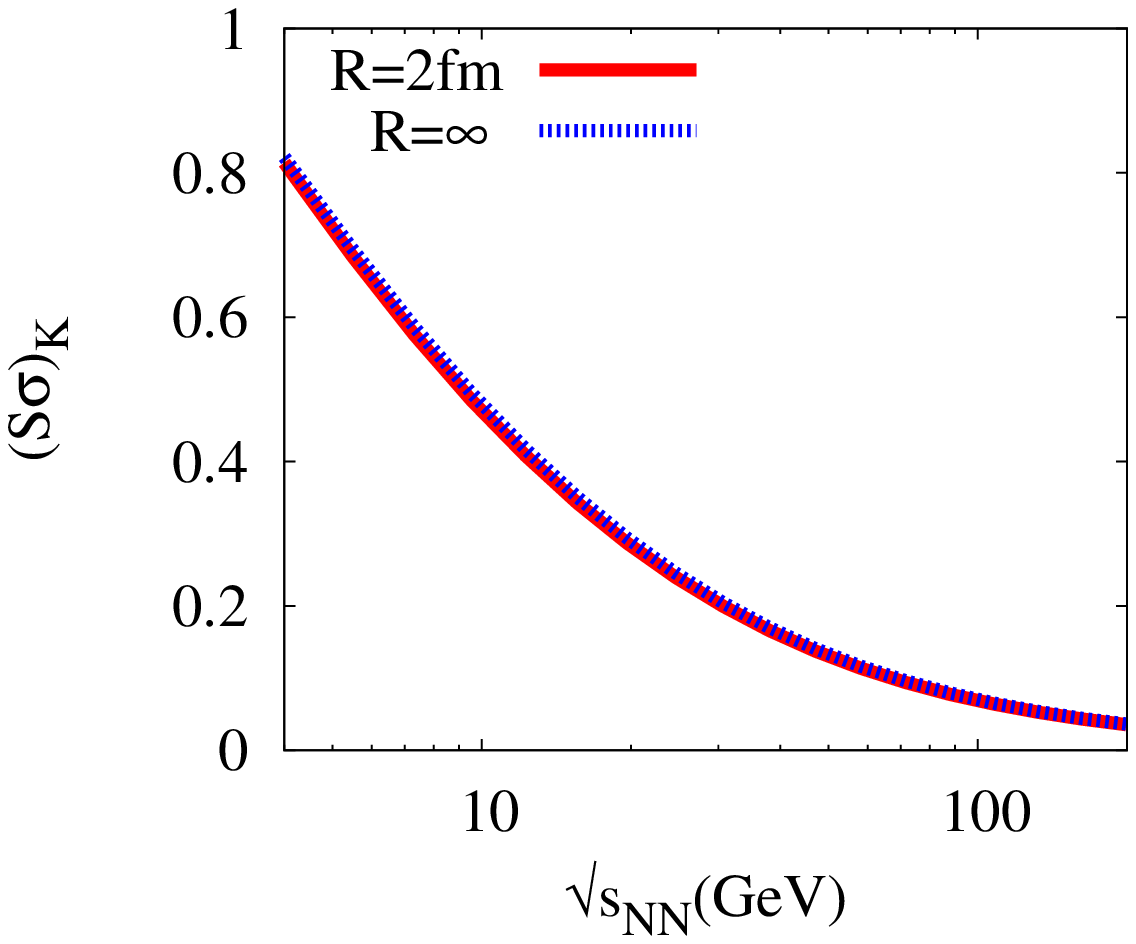}
               \label {ss_nk_roots}}
   \subfigure {\includegraphics[scale=0.4]{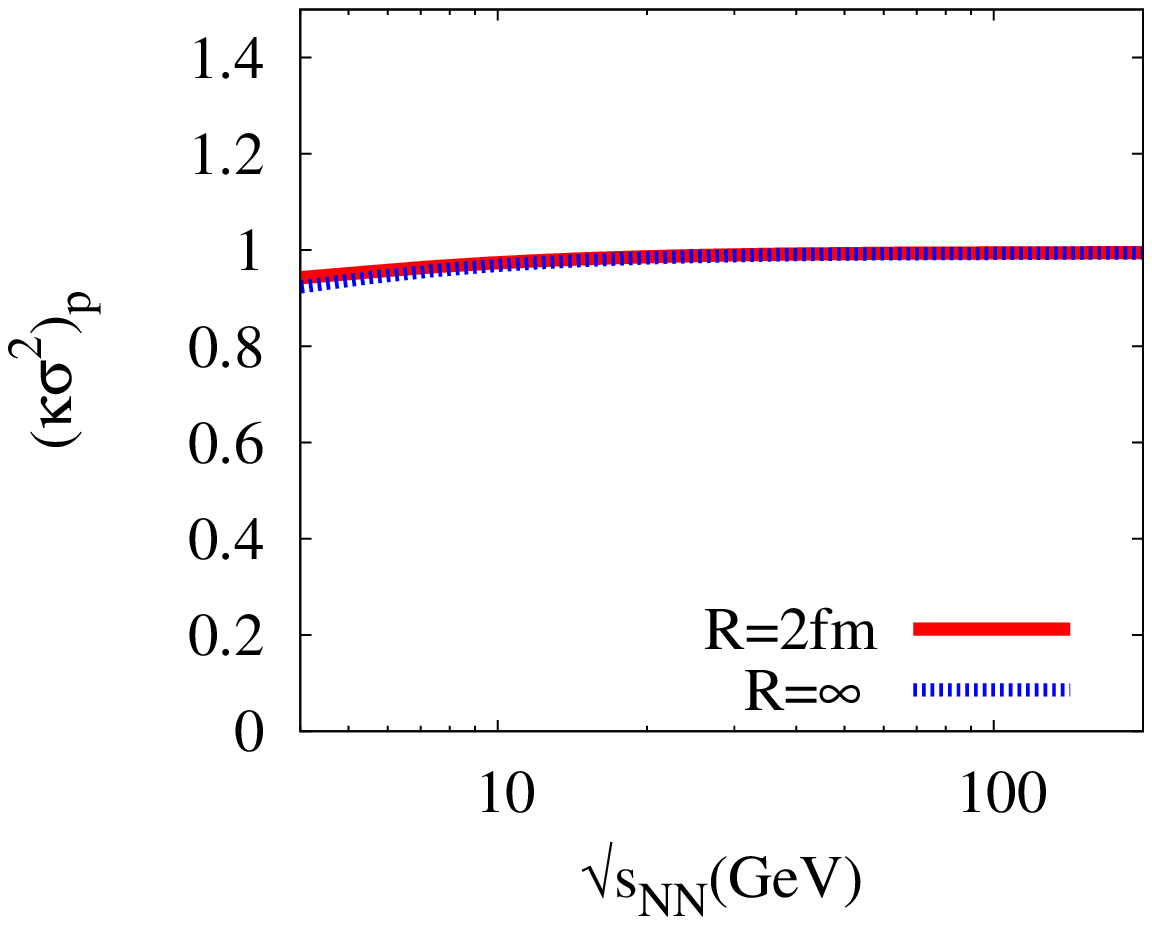}
               \label {ks2_np_roots}}
   \subfigure {\includegraphics[scale=0.4]{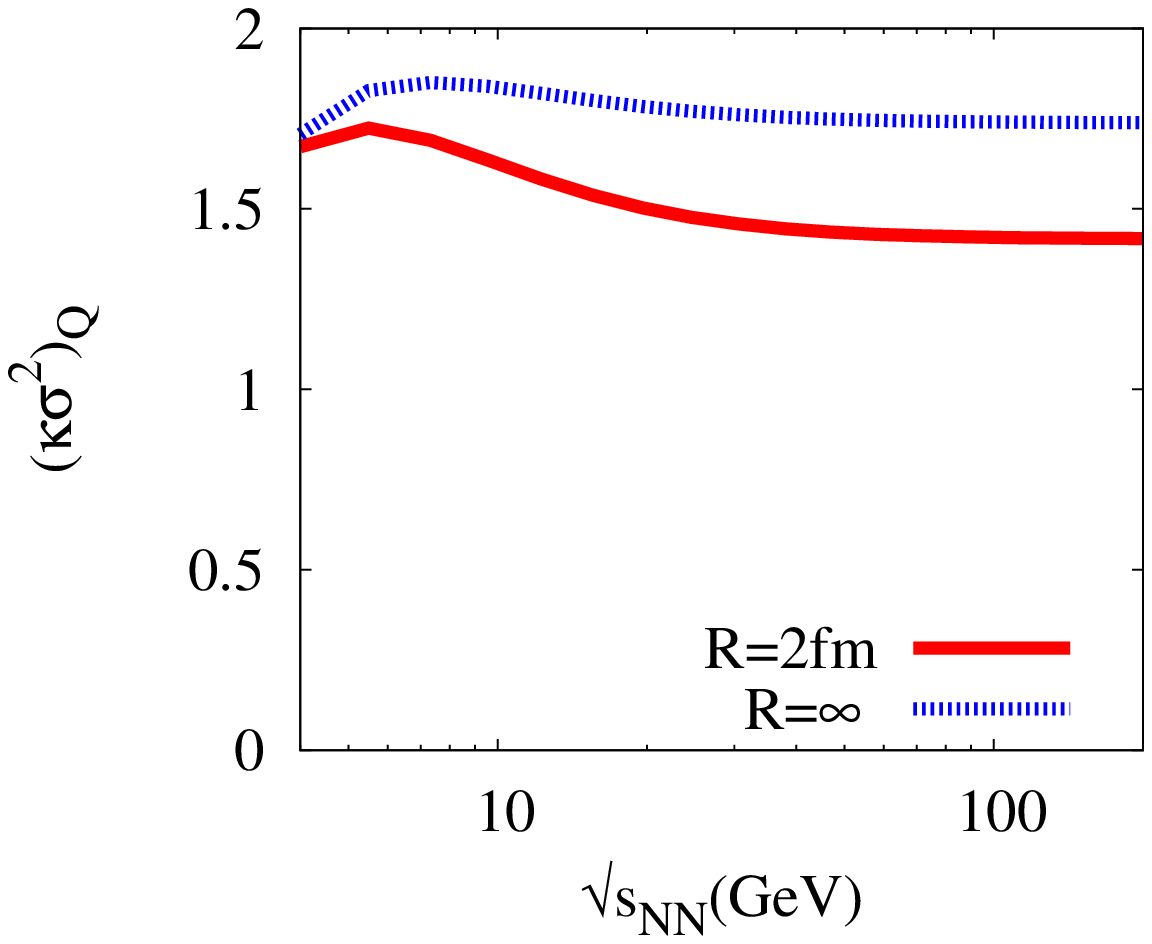}
               \label {ks2_nc_roots}}
   \subfigure {\includegraphics[scale=0.4]{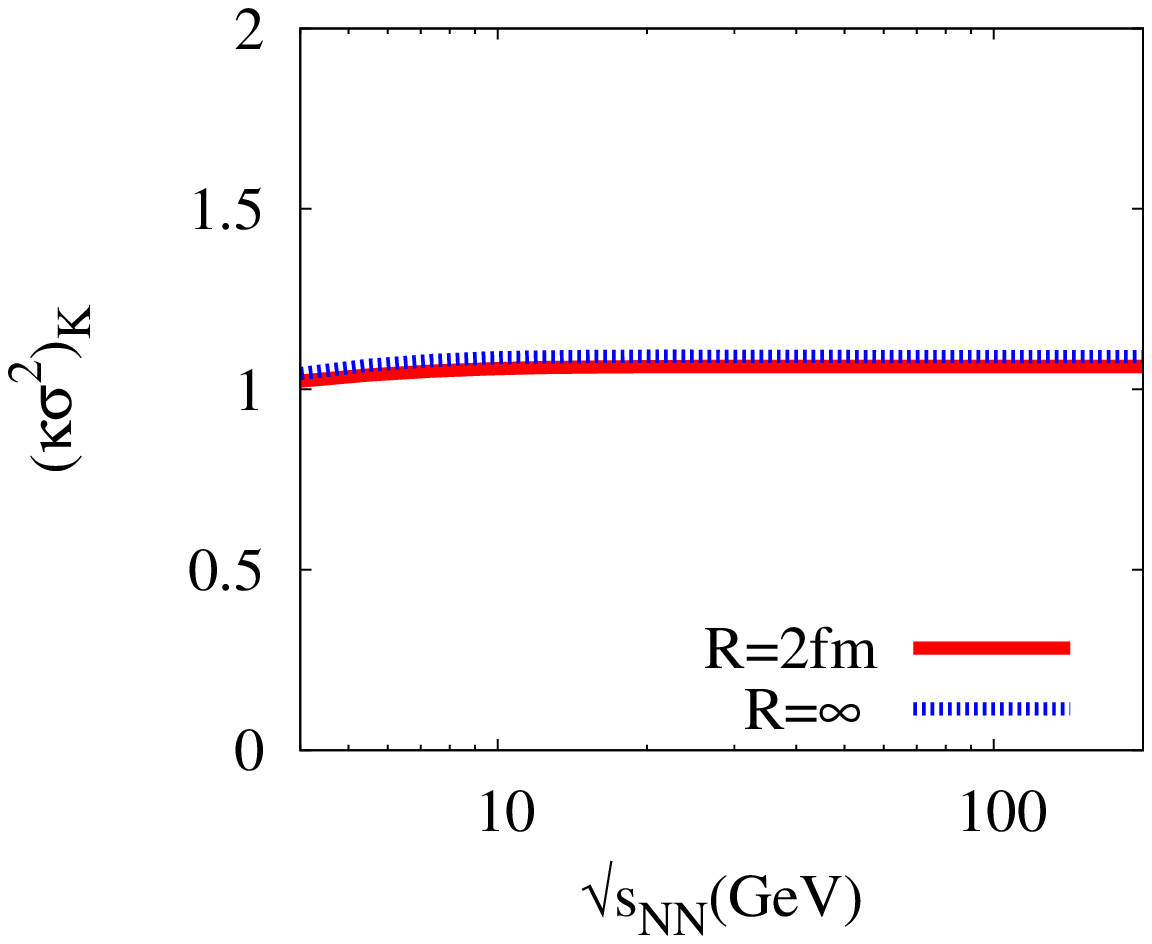}
               \label {ks2_nk_roots}}
 \caption{(Color online) Energy dependence of $\sigma^2/M$, $S\sigma$
          and $\kappa\sigma^2$  for proton (left column), net electric charge (middle column) and net kaon 
 (right column).}
 \label{fig:np_roots}
\end{figure}
This is further reflected in the observables  mean ($M$), 
variance ($\sigma$), 
skewness ($S$) and kurtosis ($\kappa$) describing the particle distribution. These
observables may be combined to relate to the ratios of susceptibilities
as,
\begin{equation}\label{moment_product}
\frac{\sigma_x^2}{M_x}= \frac{\chi_x^2}{\chi_x^1},~~~~ 
S_x\sigma_x= \frac{\chi_x^3}{\chi_x^2},~~~~ 
\kappa_x\sigma_x^2=\frac{\chi_x^4}{\chi_x^2}.
\end{equation}
These kind of ratios are important when comparing theoretical
predictions with experimental results where the volumes may not be
estimated reliably and it is expected that the volume factors in the
numerator and denominator drop out \cite{allton3}. However this
assumption is valid when the low momentum scales are less important.
The variation of these ratios for net proton, net charge and net kaon
are shown as a function of the centre of mass collision energy 
$\sqrt{s}$ in Fig. (\ref{fig:np_roots}). The parametrization of the
temperature and chemical potentials at the freeze-out conditions is
taken from Ref.~\cite{redlich}. We find that for protons and kaons the ratios
are almost independent of volume, but for the net electric charge sector
a significant volume dependence is present. While the proton and kaon
masses are in the range of $0.5 - 1$ GeV, the lowest mass in the
electric charge sector is that of the pions with is similar to
the lowest momentum scale for a finite size system. A similar effect
of low momentum scale affecting the results at cross-over has been
discussed in~\cite{sur}.
 
To summarise, we have studied the thermodynamic properties and
fluctuations of hot and dense matter in a finite volume. We have used
the HRG model for this purpose. The finite volume has a significant
effect on the thermodynamic properties and also on the fluctuations. But
for most observables the effect diminishes once they are scaled with the
respective volumes. The only significant difference even after scaling
with the volume is observed for the fluctuations for electric charge.
Given that one can relate such volume dependence on the observable
in heavy-ion collision experiments, it may be possible to extract
information about the system volume from the net electric charge fluctuations.

\section *{Acknowledgement}

This work is funded by Council for Scientific and Industrial Research
(CSIR) (Government of India), Department of Science and Technology (DST)
(Government of India) and Alexander von Humboldt (AvH) Foundation
(Germany).

\end{document}